\documentclass[letterpaper, 12pt]{article}
\usepackage[english]{babel}
\usepackage{csquotes}
\usepackage{amssymb}
\usepackage{amsmath}
\usepackage{enumitem}
\usepackage{graphicx}

\usepackage{titlesec}				

\titleformat{\section}{\large\scshape\raggedright}{}{0em}{}[\titlerule]
\titlespacing{\section}{0pt}{10pt}{10pt}

\begin{document}

\begin{center}
    \Large{On The Physical Non-Equivalence of Chiral Bases} \\ \vspace{5 mm} 
    \large{T. B. Watson and Z. E. Musielak}\\
    \small{Department of Physics, University of Texas at Arlington \\ Arlington, TX 76019, USA}
\end{center}

\begin{abstract}
In this letter we seek to redress lingering misconceptions pertaining to the physicality of the chiral phase of Dirac bi-spinor fields. Demonstrably, the most general first-order partial differential equation for spinor wavefunctions that can be obtained in Minkowski spacetime is the Dirac-like equation which leaves both the mass and chiral angles as free parameters, the so-called Chiral Dirac Equation. Previously, claims have plauged the literature which assert that any attempt to incorporate chirality by such a generalization can be trivially reduced to the case the nominal Dirac Equation. These statements are incorrect. In this letter we present a formal proof demonstrating the physical non-equivalence of particle states whose chiral angles differ, thereby demonstrating unequivocally the physicality of the chiral basis. 
\end{abstract}

\section{Introduction}

It is well-known that ad hoc addition of both scalar and pseudoscalar mass terms terms to the Dirac Lagrangian will contribute to the propagation mass of the free, bi-spinor field $\psi$ [10-12]. That the addition of such terms constitute the most general Poincar\'e-invariant generalizations of Dirac's original first-order equation is readily demonstrable [13-15]. The solutions to this generalized class of equations may be identified as particle states of definite mass and spin which transform as induced, irreducible representations of the group of proper, orthochonous Lorentz transformations and space-time translations. 

Historically, the observation that any one of these equations is connected to the Dirac equation via a unitary transformation has led to the misguided assumption of the physical equivalence of the corresponding irreducible representations. What is lost in this mathematical reduction is the subtlety of discrete physical symmetries and the

In this Letter, we present a formal proof 
that there is no unitary transformation that connects the set of positive-energy, parity-eigenstate solutions of the CDE to the set of positive-energy solutions of the DE.  Thus, the results of this Letter confirm that the CDE is the most general equation for spinor wavefunctions that can be constructed in 
Minkowski spacetime, whose metric is invariant with respect to the Poincar\'e group. In the light of these results, the Letter also discusses the implications QFT and modern physics. 

\section{Origin of the Chiral Dirac Equation}

It is useful for our discussion to define the Chiral Dirac Equation as the most general can be written in the following form
\begin{equation}
\label{eqn:CDE}
(i \gamma^{\mu} \partial_\mu - m e^{-i \alpha \gamma^5})\psi = 0\ ,
\end{equation}
where $\alpha$ is the chiral angle, which is here a necessary degree of freedom;
the equation reduces to the DE when $\alpha = 0$.  The CDE was originally 
derived by using several different methods [13-15], which are now briefly 
described.

One of them required using irreducible representations (irreps) of the Poincar\'e
group $\mathcal {P}\ =\ SO (3,1) \otimes_s T (3+1)$, with $SO(3,1)$ being a 
non-invariant Lorentz group of rotations and boosts, and $T(3+1)$ an invariant 
subgroup of spacetime translations. The condition that the Dirac spinor wavefunction 
transforms as one of the irreps of $T(3+1)\subset \mathcal {P}$ extended by parity 
is used to define elementary particles of the theory [16,17].  Since this method of 
deriving the CDE is based explicitly on the symmetries of Minkowski spacetime, we 
may be certain the resulting equation is Poincar\'e invariant.  Furthermore, the CDE 
correctly accounts for all four sets of states (spins up and down, matter and antimatter), 
and its solutions have important physical implications [13].  

The Lagrangian formalism is a powerful and independent way to derive a dynamical 
equation. The Lagrangian for the DE is very well-known and presented in textbooks 
(e.g., [18,19]) without derivation. In fact, the Lagrangian was not a part of the Dirac's 
original paper where the equation first appeared [1].  An interesting attempt to obtain 
the Dirac Lagrangian is presented and discussed in [20].  Similar approach was used 
to find the Lagrangian for the CDE [13,15].  Another method of deriving the CDE is 
based on projection operators and as expected the same CDE is obtained [13,15].  
Being Poincar\'e invariant, local and having its Lagrangian, the CDE possess all
required characteristics to be called the fundamental equation of physics.

The above statement has been contradicted by some previous work [10-12] in 
which attemps were made to include ad hoc pseudoscalar mass terms into the DE.  
Then, it was shown that solutions to the DE with the pseudoscalar terms reduce 
to those already known solutions to the DE without the terms.  In the following, 
we demonstrate these results are physically incorrect by presenting a proof of 
the physical inequivalence between the representations related by a chiral rotation.
 
\section{On the Distinctions of the Dirac Equation and its Chiral Forms}

Given the Dirac equation
\begin{align*}
\left(i\gamma^\mu\partial_\mu-m\right)\psi=0
    \end{align*}
We define a particle state in the chiral basis $\alpha=0$ as the positive-energy ($+E$), positive-parity solutions
\begin{align*}
    \psi^+\left(x;0\right)=u_s\left(p;0\right)e^{-ip\cdot x}
\end{align*}
and take the antiparticle spinors to be given by the negative-energy $-E$, negative-parity solutions:
\begin{align*}
    \psi^-\left(x;0\right)=v_s\left(p;0\right)e^{+ip\cdot x}
\end{align*}
Now, there will always exist a unitary transformation which allows one to transform the Dirac equation into the Eq. (\ref{eqn:CDE}), and vice versa. It is given by
\begin{align*}
    \psi\rightarrow U\psi=e^{\frac{i\alpha}{2}\gamma^5}\psi
\end{align*}
and constitutes an alternate choice of chiral basis\footnote{Crucially, $U$ is not a unitary \emph{spinor} transformation of the type $\psi \rightarrow 
V\psi$ and $\gamma^\mu \rightarrow V \gamma^\mu V^\dagger$. Rather, $U$ acts only on the composite bi-spinor and not jointly on the underlying Clifford Basis}. Applying the unitary chiral rotation operator above to our $+E$ 
Dirac solutions, we find
\begin{align*}
\psi^+\left(x;0\right)\rightarrow\psi^+\left(x;\alpha\right)&=e^{\frac{i\alpha}{2}\gamma^5}
\psi^+\left(x\right)=u_s\left(p;\alpha\right)e^{-ip\cdot x}
\end{align*}
and
\begin{align*}
\psi^-\left(x;0\right)\rightarrow\psi^-\left(x;\alpha\right)&=e^{\frac{i\alpha}{2}\gamma^5}
\psi^-\left(x,t\right)=v_s\left(p;\alpha\right)e^{+ip\cdot x}
\end{align*}
for 
\begin{align*}
u_s\left(p;\alpha\right) &\equiv \left[\cos{\frac{\alpha}{2}}u_s\left(p;0\right)+i\sin{\frac{\alpha}{2}}
v_{3-s}\left(p;0\right)\right] \\ 
v_s\left(p;\alpha\right) &\equiv \left[\cos{\frac{\alpha}{2}}v_s\left(p;0\right)+i\sin{\frac{\alpha}{2}}
u_{3-s}\left(p;0\right)\right]
\end{align*}
These chiral-rotated states now satisfy Eq. (\ref{eqn:CDE}). In investigating the parity properties of these representations, recall that the intrinsic parity of a fermionic 
field is defined by the action of the parity operator ($\hat{P}=\lambda \gamma^0$ for $\lambda\in \{\pm 1,\pm i\}$) on the at-rest solutions. Because a chiral rotation is not a unitary spin transformation, $\hat{P}$ remains invariant. Therefore, 
in the rest frame the CDE equations reduce to:
\begin{align*}
\hat{P}u_s\left(0;\alpha\right)=+e^{-i\alpha\gamma^5}u_s\left(0;\alpha\right) \\
\hat{P}v_s\left(0;\alpha\right)=-e^{-i\alpha\gamma^5}v_s\left(0;\alpha\right)
\end{align*}
and so $u_s\left(p;\alpha\right)$ and $v_s\left(p;\alpha\right)$ are not generally eigenstates of parity. 
This is significant as it implies a unitary chiral rotation will carry $+E$, parity-eigenstate solutions of 
the Dirac equation into $+E$ solutions of the more general Eq. (\ref{eqn:CDE}) with mixed parity. 

We are, of course, free to construct states with well-defined intrinsic parity in the $\alpha\neq0$ basis. 
These are not difficult to obtain and are given simply in the rest frame by:
\begin{align*}
\psi^A\left(x;\alpha\right)&=e^{-i\alpha\gamma^5/2}u_s\left(0;\alpha\right)e^{-imt}=u_s\left(0;0\right)e^{-imt} \\
\psi^B\left(x;\alpha\right)&=e^{-i\alpha\gamma^5/2}v_s\left(0;\alpha\right)e^{+imt}=v_s\left(0;0\right)e^{+imt}
\end{align*}

\begin{figure}
\label{fig:Energy_Parity_Diagram}
\centering
\includegraphics[width=0.75\textwidth]{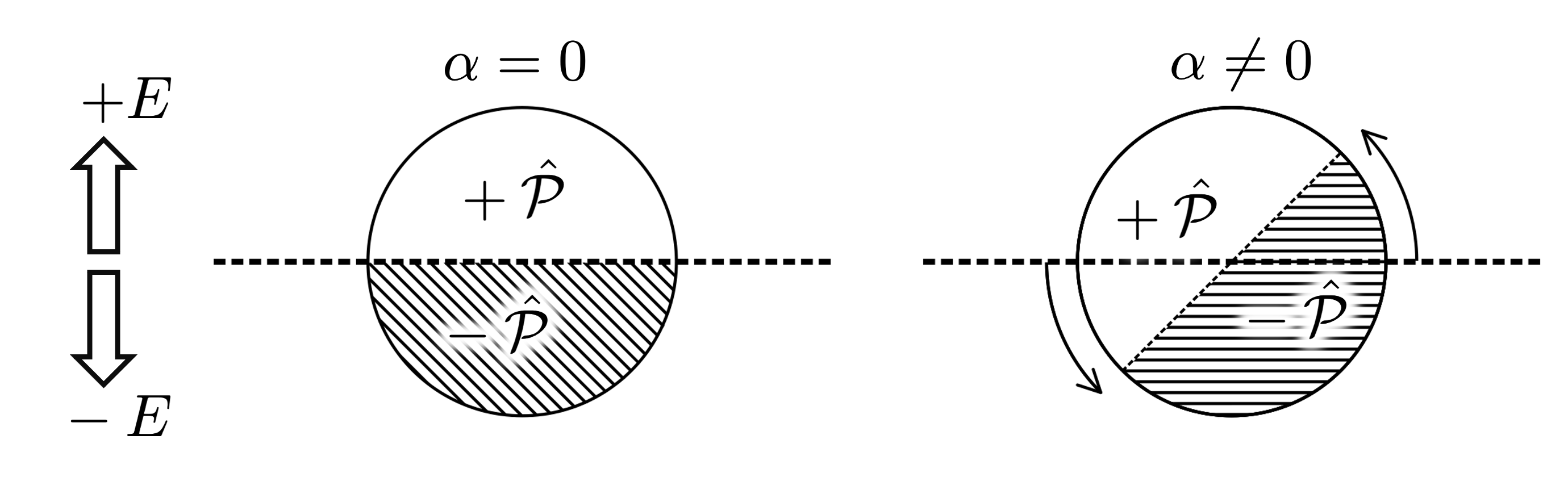}

\caption{A diagrammatic representation of the effects of a unitary chiral rotations 
on the set of solutions to the Dirac equation ($\alpha=0$) and the general form 
($\alpha\neq 0$) presented in the text. The bijectivity of the unitary transformation results in positive energy solutions of mixed parity in the general case. The Dirac equation is thereby seen to be a special case in which eigenstates of intrinsic 
parity and energy align. (Note the antiparticle diagram follows via a reversal of the 
parity signs.}
\end{figure}

However, as we are currently working in the basis where $\alpha\neq0$, these particle/anti-particle states 
are related to the solutions in the basis where $\alpha=0$ via the inverse transformation $U^\dag=
\exp{\left(-i\alpha\gamma^5/2\right)}$. Hence the parity eigenstates $\psi^A\left(x;\alpha\right)$ and 
$\psi^B\left(x;\alpha\right)$ correspond in the $\alpha=0$ basis to the at-rest solutions given by:
\begin{align*}
\psi^A\left(x;\alpha\right)\rightarrow\psi^A\left(x;0\right)&=e^{-i\alpha\gamma^5/2}\psi^A\left(x;
\alpha\right)=e^{-i\alpha\gamma^5}u_s\left(0;0\right)e^{-imt}\\ \qquad &=\left(\cos{\frac{\alpha}{2}}
u_s\left(0;0\right)-i\sin{\frac{\alpha}{2}}v_{3-s}\left(0;0\right)\right)e^{-imt}\\
\psi^B\left(x;\alpha\right)\rightarrow\psi^B\left(x;0\right)&=e^{-i\alpha\gamma^5/2}\psi^B\left(x;
\alpha\right)=e^{-i\alpha\gamma^5}v_s\left(0;0\right)e^{+imt}\\ \qquad &=\left(\cos{\frac{\alpha}{2}}
v_s\left(0;0\right)-i\sin{\frac{\alpha}{2}}u_{3-s}\left(0;0\right)\right)e^{+imt}
\end{align*}

Note the appearance of $v_{3-s}\left(p;0\right)e^{-imt}$ and $u_{3-s}\left(p;0\right)e^{+imt}$. These 
are states of positive energy and negative parity, and negative energy and positive parity respectively. Thus, we have proven the following: 

\emph{No unitary transformation exists which connects the set of positive-energy, parity-eigenstate solutions 
of the Chiral Dirac Equation to the set of positive-energy solutions of the Dirac Equation.}\footnote{The only 
exception to this is the trivial case of the identity matrix when $\alpha=0$.}

The implications of this statement are physically meaningful when taken in conjunction with the requirement that any well-defined QFT should possess only states with bounded energies. It is then the combination of these facts that amount to the physical inequivalence of the representations. This is the central point which some authors [10-12] have failed to properly appreciate, and is the reason we argue the CDE constitutes a non-trivial generalization of the DE whose solutions form physically inequivalent irreducible representations of the Poincar\'{e} group. 

To summarize: The formulation of any QFT for spin-half fields constrained by the continuous symmetries of flat space-times must begin with the chiral-free form of the Dirac Equation (CDE). It is only by the assumption of additional discrete symmetries that one may proceed to more specific forms (e.g., the nominal Dirac Equation). What previous authors have failed to appreciate is that the assumption of discrete symmetries is a physically-meaningful statement and so must result in a physically distinct theory. It is our hope that the above proof will serve to definitively clarify this misconception and facilitate a more precise dialogue when exploring the larger permissible parameter space. 

\bigskip\noindent
{\bf Acknowledgments}
We would like to thank Ben Jones for bringing to our attention several papers 
on the role of chirality in QFT and helpful discussions.
%

%--------------------------------References------------------------


\begin{thebibliography}{qqq}

\bibitem{1} P.A.M. Dirac, Proc. Royal Soc. London 117 (1928) 610
\bibitem{2} I. Sogami, Prog. Theor. Phys. 66 (1981) 303
\bibitem{3} S.I. Kruglov, arXiv:hep-ph / 0507027v2 23 June 2006
\bibitem{4} E. Marsh, Y. Narita, Front. Phys. 3 (2015) Article 82
\bibitem{5} A.O. Barut, P. Cordero, G.C. Ghirardi, Phys. Rev. 182 (1969) 1844
\bibitem{6} A.O. Barut, The mass of muon, Phys. Lett. 73B (1978) 310
\bibitem{7} S.I. Kruglov, Elect. J. Theor. Phys. 3 (2006) 11
\bibitem{8} S.I. Kruglov, Phys. Let. B  718 (2012) 228
\bibitem{9} K. Nozari, Chaos, Solitons \& Fractals 32 (2007) 302
\bibitem{10} D. Leiter, G. Szamosi, Let. Nuovo Cim. 5 (1972) 814
\bibitem{11} A. Da Silveirada, Lett. Nuovo Cim. 15 (1976) 228
\bibitem{12} M. Trzetrzelewski, arXiv preprint (2011) arXiv: 1101.3899
\bibitem{13} T.B. Watson, Z.E. Musielak, Int. J. Mod. Phys. A 35 (2020) 2050189
\bibitem{14} T.B. Watson, Z.E. Musielak, Symmetry 13 (2021) 1608
\bibitem{15} T.B. Watson, PhD Thesis, The Univeristy of Texas at Arlington, 
                    August 2022
\bibitem{16} E.P. Wigner, On unitary representations of the inhomogeneous Lorentz group, 
                   Ann. Math. 40 (1939) 149
\bibitem{17} Y.S. Kim and M.E. Noz, Theory and Applications of the 
                  Poincar\'e Group, Reidel, Dordrecht, 1986
\bibitem{18} L.W. Ryder, Quantum Field Theory, Cambridge University Press,
                  Cambridge, 1985
\bibitem{19} P.H. Frampton, Gauge Field Theories, John Wiley \& Sons, Inc., 
                  New York, 2000
\bibitem{20} N.A. Daughty, Lagrangian Interactions, Addison-Wesley Publ. Comp., Inc.,
                   Sydney, 1990
\end{thebibliography}
\end{document}